\documentclass[12pt]{article}
\usepackage{graphicx}

\begin{document}

\title{Noise reduction caused by eavesdropping on six-state quantum key distribution over collective-noise channel}

\author{Hiroo Azuma\thanks{On leave from Nisshin-scientia Co., Ltd.,
8F Omori Belport B, 6-26-2 MinamiOhi,
Shinagawa-ku, Tokyo 140-0013, Japan.
Email: hiroo.azuma@m3.dion.ne.jp, zuma@nii.ac.jp}
\\
\\
{\small Global Research Center for Quantum Information Science,}\\
{\small National Institute of Informatics,}\\
{\small 2-1-2 Hitotsubashi, Chiyoda-ku, Tokyo 101-8430, Japan}\\
}

\date{\today}

\maketitle

\begin{abstract}
In this paper,
we show that there are instances where eavesdropping causes noise reduction for a quantum key distribution (QKD) protocol.
To witness these phenomena,
we investigate a fault-tolerant six-state QKD protocol over a collective unitary noise channel.
In this protocol, legitimate users send and receive two-qubit states that belong to the noiseless subspace being robust against collective unitary errors.
We examine eavesdropper's intercept/resend and entangling probe attacks on this protocol.
In general, the collective unitary noises lessen the probability that legitimate users share a random bit with the QKD protocol.
However, we show that eavesdropping enlarges that probability in some specific scenarios
although the effects of the collective unitary noise channel are strong enough.
These phenomena make the legitimate users difficult to distinguish between noises and eavesdropper's malicious acts by monitoring the probability
that they share the same random key.
\end{abstract}

\section{\label{section-introduction}Introduction}
The six-state quantum key distribution (QKD) protocol
\cite{Bruss1998,Bechmann-Pasquinucci1999}
is a natural extension of the well-known BB84 four-state scheme
which was proposed by Bennett and Brassard in 1984 \cite{Bennett1984}.
The six-state protocol uses three bases rather than two ones
that are utilized by the BB84 scheme.
Although the six-state QKD protocol is regarded as less practical than the BB84 protocol,
it has theoretically interesting features so many researchers have investigated it.

Because the BB84 scheme has a long tradition,
its properties have been studied eagerly and vastly \cite{Bennett1992a,Bennett1992b}.
In particular,
its practical aspects have been investigated very much \cite{Lutkenhaus1999,Lutkenhaus2000,Gottesman2004,Scarani2009,Adachi2007,Gobby2004}.
The BB84 scheme was proven unconditionally secure against an enemy who was able to mount arbitrary attacks permitted by quantum mechanics
\cite{Lo1999,Mayers2001,Biham2006,Shor2000}.

Because the six-state protocol is a direct descendant of the BB84 protocol,
it inherits many qualities of the BB84.
The six-state protocol has been studied already by many researchers.
Bru{\ss} examined the security of the six-state protocol against eavesdropping on a single qubit \cite{Bruss1998}.
Bechmann-Pasquinucci and Gisin investigated in-coherent and coherent attacks on the six-state protocol \cite{Bechmann-Pasquinucci1999}.
The unconditional security of the six-state protocol has been proved already \cite{Lo2001}.
Information-theoretic security proof for the six-state protocol
with one-way error correction and privacy amplification was presented \cite{Renner2005}.
Intercept/resend attacks on the six-state protocol over noisy channels were examined \cite{Garapo2016}.
The security proof of the six-state protocol with threshold detectors was investigated for practical purposes \cite{Kato2016}.

To increase the practicality of the QKD protocol,
the effect of a noisy channel is a serious problem that we must overcome.
The singlet state for two qubits remains unchanged under an independent unitary noise for each qubit.
Thus, some researchers studied modifications of the BB92 and BB84 schemes by using a noiseless subspace that includes the singlet state
\cite{Zanardi1997,Kwiat2000,Boileau2004,Wang2005}.

In the current paper, we investigate the security of a fault-tolerant six-state QKD protocol over a collective unitary noise channel
against intercept/resend and entangling probe attacks.
Moreover, we show that there are some instances where these malicious acts of the eavesdropper reduce the noises
and enlarge the probability that legitimate users share a random key correctly.

As mentioned above, the singlet state is invariant under the collective unitary noise.
Taking advantage of this property,
we consider a deformed six-state QKD protocol whose six states the legitimate users send and receive belong to the noiseless subspace including the singlet state.
Thanks to this improvement, the protocol becomes robust against phase errors that are caused by the unitary noise.

We examine the security of this improved QKD protocol against eavesdropping, that is, the intercept/resend and entangling probe attacks.
In general, the unitary noise lessens the probability that Alice and Bob obtain the same random bit with the protocol.
However, we show that Eve's malicious act can make that probability increase under specific conditions
although the protocol is suffering from the unitary noise.

Because of this trouble, if Alice and Bob try to detect Eve's malicious acts by comparing their random bit strings,
they cannot distinguish bit errors caused by bit flipping of the unitary noise from those given rise to by Eve's eavesdropping.
Thus, Eve can pretend that her disturbance is noise induced by an actual channel.

This paper is organized as follows.
In Sec.~\ref{section-review-six-state-protocol-collective-unitary-noise-channel},
we review the six-state QKD protocol and the collective unitary noise channel.
In Sec.~\ref{section-noiseless-subspace-improved-six-state-protocol},
we modify the six-state protocol by letting Alice and Bob transmit quantum codes defined in the noiseless subspace.
In Sec.~\ref{section-efficiency-without-eavesdropping},
we evaluate the probability that Alice and Bob share the same random bit under the collective unitary noise channel but Eve does not interfere with the protocol.
In Sec.~\ref{section-intercept-resend-attack}, we examine the security of the improved protocol against Eve's intercept/resend attack.
In Sec.~\ref{section-entangling-probe-attack},
we investigate the security of the improved protocol against Eve's entangling probe attack.
In Sec.~\ref{section-discussions},
we give brief discussions.
In Appendices, we show some useful equations and mathematical expressions utilized in the main text.
In Supplemental Material, we show some mathematical derivations of equations given in the main text.

\section{\label{section-review-six-state-protocol-collective-unitary-noise-channel}
Reviews of the six-state QKD protocol and the collective unitary noise channel}
The six-state protocol is a natural extended version of the BB84 scheme.
In the BB84 scheme, Alice and Bob transmit the following four states at random:
\begin{eqnarray}
&&
|0\rangle,
\quad
|1\rangle, \nonumber \\
&&
|\pm\rangle
=
(1/\sqrt{2})(|0\rangle\pm|1\rangle).
\end{eqnarray}
In addition to these states,
the six-state QKD protocol utilizes the following two states:
\begin{equation}
|\pm i\rangle
=
(1/\sqrt{2})(|0\rangle\pm i|1\rangle).
\end{equation}
We draw attention to the fact that these six states are eigenvectors of $\sigma_{z}$, $\sigma_{x}$, and $\sigma_{y}$.

The collective unitary noise channel is defined as follows \cite{Zanardi1997}.
If we transmit quantum states
$|0\rangle=(1,0)^{\mbox{\scriptsize T}}$ and $|1\rangle=(0,1)^{\mbox{\scriptsize T}}$ through this channel,
it transforms them as
\begin{equation}
|k\rangle
\rightarrow
U|k\rangle
\quad
\mbox{for $k\in\{0,1\}$},
\label{collective-unitay-noise-0}
\end{equation}
where
\begin{equation}
U
=
\left(
\begin{array}{cc}
\cos\theta & -e^{i(\delta-\phi)}\sin\theta \\
e^{i\phi}\sin\theta & e^{i\delta}\cos\theta \\
\end{array}
\right).
\label{collective-unitay-noise-1}
\end{equation}
If we transmit $N$ qubits via the collective unitary noise channel,
it gives rise to errors of the qubits in the form,
\begin{equation}
\rho_{N}
\rightarrow
(U)^{\otimes N}\rho_{N}(U^{\dagger})^{\otimes N}.
\label{collective-unitay-noise-2}
\end{equation}
As shown in the above equation,
the collective unitary noise channel applies the unitary operator $U$ to each qubit independently.

The simplest method for removing errors caused by the collective unitary noise is to apply $U^{-1}$
to each qubit that suffers from the unitary transformation $U$.
However, it is difficult to realize this method practically.
In general,
three parameters $\delta$, $\phi$, and $\theta$ determine the unitary operator $U$ and they vary at random as time proceeds.
For example, the optical fibre gives rise to the random unitary rotation $U$.

Here, we describe the typical timescale during which variations of the three parameters develop
into thermal and mechanical fluctuations as $\tau_{\mbox{\scriptsize fluc}}$.
If $\tau_{\mbox{\scriptsize fluc}}$ is longer than the time taken by the qubit to travel from Alice to Bob,
the noise is well approximated by Eqs.~(\ref{collective-unitay-noise-0}), (\ref{collective-unitay-noise-1}), and (\ref{collective-unitay-noise-2})
and
the parameters change considerably during the transmission.
Thus, because legitimate users cannot predict variations of the parameters,
they cannot make unitary compensation by applying $U^{-1}$ to each qubit.

In this paper, we assume that Alice, Bob, and Eve cannot keep up the random variations of $\delta$, $\phi$, and $\theta$, perfectly.
We consider that observation of the parameters $\delta$, $\phi$, and $\theta$ before every transmission of the single qubit is very cumbersome and do not regard it as practical.
Hence, due to these situations, Alice and Bob must remove errors in a different way than applying $U^{-1}$ to each qubit.

\section{\label{section-noiseless-subspace-improved-six-state-protocol}
The noiseless subspace and an improved six-state protocol}
An orthogonal basis of a Hilbert space of two qubits is given by $\{|00\rangle,|01\rangle,|10\rangle,|11\rangle\}$.
Here, we examine how the collective unitary noise channel defined by Eqs.~(\ref{collective-unitay-noise-0}), (\ref{collective-unitay-noise-1}), and (\ref{collective-unitay-noise-2})
transforms $|01\rangle$ and $|10\rangle$,
\begin{eqnarray}
U|0\rangle\otimes U|1\rangle
&=&
e^{i\delta}
(
-e^{-i\phi}\sin\theta\cos\theta|00\rangle
+\cos^{2}\theta|01\rangle
-\sin^{2}\theta|10\rangle
+e^{i\phi}\sin\theta\cos\theta|11\rangle
), \nonumber \\
U|1\rangle\otimes U|0\rangle
&=&
e^{i\delta}
(
-e^{-i\phi}\sin\theta\cos\theta|00\rangle
-\sin^{2}\theta|01\rangle
+\cos^{2}\theta|10\rangle
+e^{i\phi}\sin\theta\cos\theta|11\rangle
). \nonumber\\
\end{eqnarray}
Looking at the above equation, we note that phase shift errors for $\delta$ and $\phi$ do not occur and bit flip errors for $\theta$ leave the state
in the subspace spanned by $\{|01\rangle,|10\rangle\}$.
In particular, the singlet $|\Psi^{-}\rangle=(1/\sqrt{2})(|01\rangle-|10\rangle)$ is invariant under the transformation of $U\otimes U$.
Thus, we can regard the subspace spanned by $\{|01\rangle,|10\rangle\}$ as a noiseless subspace \cite{Kwiat2000}.

Hence, using six states defined in the noiseless subspace for the six-state QKD protocol,
we can expect that it is robust against the phase shift errors caused by the collective unitary noise channel.
According to this idea, we improve the six-state protocol as follows.
(Modifications of BB92 and BB84 schemes with this plan were investigated \cite{Boileau2004,Wang2005}.)

First of all, we define a unitary transformation $V_{12}$ for two qubits in the form,
\begin{eqnarray}
&&
V_{12}|0\rangle_{1}|0\rangle_{2}=|0\rangle_{1}|1\rangle_{2},
\quad
V_{12}|0\rangle_{1}|1\rangle_{2}=|0\rangle_{1}|0\rangle_{2}, \nonumber \\
&&
V_{12}|1\rangle_{1}|0\rangle_{2}=|1\rangle_{1}|0\rangle_{2},
\quad
V_{12}|1\rangle_{1}|1\rangle_{2}=|1\rangle_{1}|1\rangle_{2}.
\end{eqnarray}
Further, we prepare the following six states:
\begin{eqnarray}
&&
V_{12}|0\rangle_{1}|0\rangle_{2}=|01\rangle_{12},
\quad
V_{12}|1\rangle_{1}|0\rangle_{2}=|10\rangle_{12}, \nonumber \\
&&
V_{12}|+\rangle_{1}|0\rangle_{2}=|\Psi^{+}\rangle_{12},
\quad
V_{12}|-\rangle_{1}|0\rangle_{2}=|\Psi^{-}\rangle_{12}, \nonumber \\
&&
V_{12}|i+\rangle_{1}|0\rangle_{2}=|i\Psi^{+}\rangle_{12},
\quad
V_{12}|i-\rangle_{1}|0\rangle_{2}=|i\Psi^{-}\rangle_{12},
\end{eqnarray}
where
\begin{eqnarray}
|\Psi^{\pm}\rangle_{12}
&=&
(1/\sqrt{2})(|01\rangle_{12}\pm|10\rangle_{12}), \nonumber \\
|i\Psi^{\pm}\rangle_{12}
&=&
(1/\sqrt{2})(|01\rangle_{12}\pm i|10\rangle_{12}).
\end{eqnarray}
To compute a parity bit,
we
define a unitary transformation $V_{\mbox{\scriptsize parity}}$ for three qubits as
\begin{equation}
V_{\mbox{\scriptsize parity}}|j\rangle_{1}|k\rangle_{2}|l\rangle_{3}
=
|j\rangle_{1}|k\rangle_{2}|j\oplus k\oplus l\rangle_{3}
\quad
\mbox{for $j,k,l\in\{0,1\}$}.
\end{equation}

An improved protocol is given as follows.
\begin{enumerate}
\item
Alice chooses a basis from the $\sigma_{x}$, $\sigma_{y}$, and $\sigma_{z}$ bases at random.
If she chooses the $\sigma_{x}$ basis, she picks a state from $\{|+\rangle,|-\rangle\}$ at random.
If she chooses the $\sigma_{y}$ basis, she picks a state from $\{|+i\rangle,|-i\rangle\}$ at random.
If she chooses the $\sigma_{z}$ basis, she picks a state from $\{|0\rangle,|1\rangle\}$ at random.
We describe the state Alice selects as $|\varphi\rangle$ and we make it a state of the first qubit.
Next, Alice attaches the second qubit $|0\rangle_{2}$ to $|\varphi\rangle_{1}$,
applies the unitary transformation $V_{12}$ to them, and obtains $|\psi\rangle_{12}=V_{12}|\varphi\rangle_{1}|0\rangle_{2}$.
Therefore, her state $|\psi\rangle_{12}$
can be any one of $\{|01\rangle_{12},|10\rangle_{12},|\Psi^{\pm}\rangle_{12},|i\Psi^{\pm}\rangle_{12}\}$.
She sends this state to Bob via the collective unitary noise channel.
\item
Receiving $|\psi'\rangle_{12}=U\otimes U|\psi\rangle_{12}$,
Bob attaches an auxiliary qubit $|0\rangle_{3}$ to it
and applies $V_{\mbox{\scriptsize parity}}$ to the three qubits.
Bob observes the third qubit with the $\sigma_{z}$ basis.
If Bob obtains $|0\rangle_{3}$, he judges that the first and second qubits lie outside the noiseless subspace and discards them.
If Bob obtains $|1\rangle_{3}$, he judges that the first and second qubits belong to the noiseless subspace and applies $V_{12}^{-1}$ to them.
Next, Bob chooses a basis from the $\sigma_{x}$, $\sigma_{y}$, and $\sigma_{z}$ bases at random
and observes the first qubit with it.
\item
Alice discloses with which basis, $\sigma_{x}$, $\sigma_{y}$, or $\sigma_{z}$, she encodes the first qubit in public via the classical channel.
\item
After the legitimate users repeat the above process,
Bob discloses in which event Alice's basis for encoding and Bob's basis for observation correspond to each other in public via the classical channel.
\item
Using events where both their bases correspond to each other,
Alice and Bob share random bits.
\end{enumerate}

\section{\label{section-efficiency-without-eavesdropping}
The probability that Alice and Bob share a random bit via the collective unitary noise channel without suffering from eavesdropping}
Even if Eve does not interfere with the transmission,
the probability that Alice and Bob share a random bit correctly varies because of the effects caused by the collective unitary noise channel.
First of all, we consider a scenario where Alice and Bob carry out the original six-state protocol.
If Alice's basis for encoding and Bob's basis for observation correspond to each other,
the probability that they share the same random bit is given by
\begin{equation}
P_{\mbox{\scriptsize original}}
=
\frac{1}{6}
[3+\cos\delta+2\cos(\frac{\delta^{2}}{4})\cos(2\theta)].
\end{equation}
Thus, if we do not utilize the noiseless subspace,
the probability that Alice and Bob obtain the same random bit depends on $\theta$ and $\delta$ but not on $\phi$.

Next, we estimate the probability that Alice and Bob share the same random bit when their bases for encoding and observation correspond to each other
under the improved six-state protocol that uses the noiseless subspace.
We can compute this probability as follows.
First, we describe the six states as column vectors in the form,
\begin{eqnarray}
|\psi_{1}\rangle_{12}
&=&
|01\rangle_{12}
=
(0,1,0,0)^{\mbox{\scriptsize T}}, \nonumber \\
|\psi_{2}\rangle_{12}
&=&
|10\rangle_{12}
=
(0,0,1,0)^{\mbox{\scriptsize T}}, \nonumber \\
|\psi_{3}\rangle_{12}
&=&
|\Psi^{+}\rangle_{12}
=
(1/\sqrt{2})(0,1,1,0)^{\mbox{\scriptsize T}}, \nonumber \\
|\psi_{4}\rangle_{12}
&=&
|\Psi^{-}\rangle_{12}
=
(1/\sqrt{2})(0,1,-1,0)^{\mbox{\scriptsize T}}, \nonumber \\
|\psi_{5}\rangle_{12}
&=&
|i\Psi^{+}\rangle_{12}
=
(1/\sqrt{2})(0,1,i,0)^{\mbox{\scriptsize T}}, \nonumber \\
|\psi_{6}\rangle_{12}
&=&
|i\Psi^{-}\rangle_{12}
=
(1/\sqrt{2})(0,1,-i,0)^{\mbox{\scriptsize T}}.
\end{eqnarray}
Second, we define a projection onto the noiseless subspace as
\begin{equation}
\Pi^{-}
=
\left(
\begin{array}{cccc}
0 & 0 & 0 & 0 \\
0 & 1 & 0 & 0 \\
0 & 0 & 1 & 0 \\
0 & 0 & 0 & 0 \\
\end{array}
\right).
\end{equation}
Then, we can compute the probability that Alice and Bob share the same random bit
on condition that their bases for encoding and observation correspond to each other as
\begin{eqnarray}
P_{\mbox{\scriptsize noiseless subspace}}
&=&
\frac{1}{6}\sum_{i=1}^{6}
|_{12}\langle \psi_{i}|\Pi^{-}(U\otimes U)|\psi_{i}\rangle_{12}|^{2} \nonumber \\
&=&
\frac{1}{6}
[3+2\cos(2\theta)+\cos(4\theta)].
\label{probability-noiseless-subspace-without-eavesdropping-0}
\end{eqnarray}
As shown in Eq.~(\ref{probability-noiseless-subspace-without-eavesdropping-0}),
the probability that Alice and Bob obtain the same random bit by using the noiseless subspace depends only on $\theta$.

\begin{figure}
\begin{center}
\includegraphics{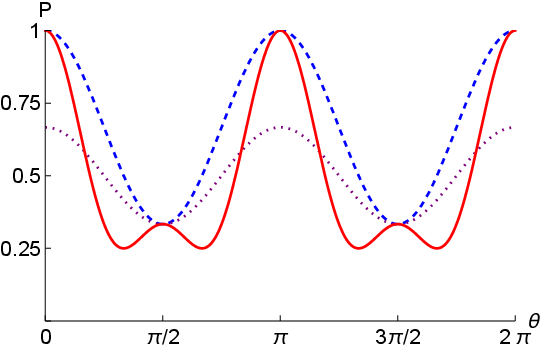}
\end{center}
\caption{Plots of $P_{\mbox{\scriptsize noiseless subspace}}$ and $P_{\mbox{\scriptsize original}}$ as functions of $\theta$.
The solid red curve represents $P_{\mbox{\scriptsize noiseless subspace}}$.
The dashed blue and dotted purple curves represent $P_{\mbox{\scriptsize original}}$ for $\delta=0$ and $\delta=\pi/2$, respectively.
The original protocol is preferable to the improved protocol under the collective unitary noise channel with $\delta=0$ for Alice and Bob.
However, if $\delta\neq 0$, for example, $\delta=\pi/2$, the improved protocol can be more advantageous to Alice and Bob
than the original protocol for a specific range of values of $\theta$.
In concrete terms, $P_{\mbox{\scriptsize noiseless subspace}}$ is larger than $P_{\mbox{\scriptsize original}}$ for $\delta=\pi/2$ and $0\leq\theta\leq\pi/6$.}
\label{Figure-01}
\end{figure}

In Fig.~\ref{Figure-01}, we plot $P_{\mbox{\scriptsize original}}$ and $P_{\mbox{\scriptsize noiseless subspace}}$ as functions of $\theta$.
Looking at these graphs, we note the following two facts.
First, if $\delta=0$, the original protocol is preferable to the improved protocol for Alice and Bob.
However, if $\delta$ is not equal to zero, for example, $\delta=\pi/2$, the improved protocol can be more beneficial to Alice and Bob than the original protocol
over $0\leq\theta\leq\pi/6$.
Thus, we can conclude that the improved protocol removes the phase errors of the collective unitary noise channel.

Second, all three plots of $P_{\mbox{\scriptsize noiseless subspace}}$ and $P_{\mbox{\scriptsize original}}$ become smaller than $1/2$ for specific ranges of values of $\theta$.
If the probability $P$ that Alice and Bob share a random bit is less than $1/2$,
some might say that Bob must inverse the value of the random bit
and make the probability equal to $1-P$.
However, Bob cannot use this technique because he does neither know nor predict the value of $\theta$.
That is, he cannot obtain the probability for each specific $\theta$ by statistical processing.
This trouble is typical and common for collective unitary noise channels.
Moreover, not only Bob but also Eve are at this disadvantage.
In this paper, we often see this trouble happen when we estimate the probabilities that legitimate users share the same random bit
and/or that eavesdropper correctly guesses at Alice's random bit.

\section{\label{section-intercept-resend-attack}Intercept/resend attacks}
In this section, we investigate which strategy is favourable to Eve if she mounts an intercept/resend attack on the improved protocol over the collective unitary noise channel.
Because the legitimate users transmit the two-qubit state over the collective unitary noise channel,
Eve can observe the two qubits using an arbitrary basis in a four-dimensional Hilbert space with a collapse of the wave function
and resend another two-qubit state according to an outcome of the observation.
However, because the degrees of freedom in this attack are very large,
it is difficult to analyse this problem.

Hence, we simplify the intercept/resend attack as follows.
First, Eve receives $(U\otimes U)|\psi_{i}\rangle$ for $i\in\{1,...,6\}$ that is a state sent by Alice and disturbed by the collective unitary noise.
(Here, although we must describe the state emitted from Alice as $|\psi_{i}\rangle_{12}$ rigorously,
we omit indices $1$ and $2$ and write it as $|\psi_{i}\rangle$ for sake of simplicity.)
Second, Eve applies the projection $\Pi^{-}$ onto the noiseless subspace to the state that she receives.
On the one hand, if the state lies outside the noiseless subspace,
Eve lets it be untouched and resends it to Bob.
On the other hand, if the state belongs to the noiseless subspace, Eve applies the optimal SU(2) rotation $V_{\mbox{\scriptsize Eve}}^{\dagger}$ to it
and observe $V_{\mbox{\scriptsize Eve}}^{\dagger}\Pi^{-}(U\otimes U)|\psi_{i}\rangle$
with a basis $\{|01\rangle,|10\rangle\}$.
Third, if Eve detects $|01\rangle$, she resends $V_{\mbox{\scriptsize Eve}}|01\rangle$ to Bob.
If Eve detects $|10\rangle$, she resends $V_{\mbox{\scriptsize Eve}}|10\rangle$ to Bob.
Further, we assume that Eve can send these states to Bob via an ideal noiseless channel.

In the above strategy, because Eve performs parity bit checking with the projection $\Pi^{-}$,
the probability that Bob receives the state lying outside the noiseless subspace is not affected by Eve's malicious acts.
Thus, Bob cannot detect Eve by monitoring the probability that the state belongs to the noiseless subspace.
Hence, this simplified intercept/resend attack is advantageous to Eve
so that this strategy is practical and worth analysing.

If we adopt the above strategy, Eve's intercept/resend attack is carried out in a two-dimensional Hilbert space spanned by $\{|01\rangle,|10\rangle\}$.
An arbitrary $2\times 2$ SU(2) matrix is given as follows \cite{Sakurai1994}:
\begin{equation}
\exp(-\frac{i}{2}\alpha\sigma_{z})\exp(-\frac{i}{2}\beta\sigma_{y})\exp(-\frac{i}{2}\gamma\sigma_{z})
=
\left(
\begin{array}{cc}
v_{11} & v_{12} \\
v_{21} & v_{22} \\
\end{array}
\right),
\label{SU2-matrix-0}
\end{equation}
where
\begin{eqnarray}
v_{11}
&=&
e^{-i(\alpha+\gamma)/2}\cos(\beta/2) \nonumber \\
v_{12}
&=&
-e^{-i(\alpha-\gamma)/2}\sin(\beta/2) \nonumber \\
v_{21}
&=&
e^{i(\alpha-\gamma)/2}\sin(\beta/2) \nonumber \\
v_{22}
&=&
e^{i(\alpha+\gamma)/2}\cos(\beta/2).
\end{eqnarray}

If Alice sends $|\psi_{i}\rangle$, the probability that Eve obtains a correct bit value that Alice chooses is given by $P_{\mbox{\scriptsize Eve},i}$ in the form,
\begin{eqnarray}
P_{\mbox{\scriptsize Eve},i}
&=&
|\langle\varphi_{01}|V_{\mbox{\scriptsize Eve}}^{\dagger}\Pi^{-}(U\otimes U)|\psi_{i}\rangle|^{2}
\quad
\mbox{for $i\in\{1,3,5\}$}, \nonumber \\
P_{\mbox{\scriptsize Eve},i}
&=&
|\langle\varphi_{10}|V_{\mbox{\scriptsize Eve}}^{\dagger}\Pi^{-}(U\otimes U)|\psi_{i}\rangle|^{2}
\quad
\mbox{for $i\in\{2,4,6\}$},
\end{eqnarray}
where
\begin{equation}
V_{\mbox{\scriptsize Eve}}=
\left(
\begin{array}{cccc}
0 & 0 & 0 & 0 \\
0 & v_{11} & v_{12} & 0 \\
0 & v_{21} & v_{22} & 0 \\
0 & 0 & 0 & 0 \\
\end{array}
\right),
\end{equation}
and
\begin{eqnarray}
|\varphi_{01}\rangle
&=&
(0,1,0,0)^{\mbox{\scriptsize T}}, \nonumber \\
|\varphi_{10}\rangle
&=&
(0,0,1,0)^{\mbox{\scriptsize T}}.
\end{eqnarray}

Thus, we can compute the probability that eve succeeds in eavesdropping $P_{\mbox{\scriptsize Eve}}$ as
\begin{eqnarray}
P_{\mbox{\scriptsize Eve}}
&=&
\frac{1}{6}\sum_{k=1}^{6}
P_{\mbox{\scriptsize Eve},k} \nonumber \\
&=&
\frac{1}{12}
[4
+\cos(\beta-2\theta)
+\cos(4\theta)
+\cos(\beta+2\theta)
+\cos\alpha\sin\beta \nonumber \\
&&
+2\cos(2\theta)\sin\alpha\sin\beta
+\cos^{2}(2\theta)(1+\cos\alpha\sin\beta)].
\label{P-Eve-intercept-resend-attack-1}
\end{eqnarray}
Here, we draw attention to the fact that $P_{\mbox{\scriptsize Eve}}$ does not depend on the parameter of the SU(2) rotation $\gamma$
given by Eq.~(\ref{SU2-matrix-0}).

\begin{figure}
\begin{center}
\includegraphics{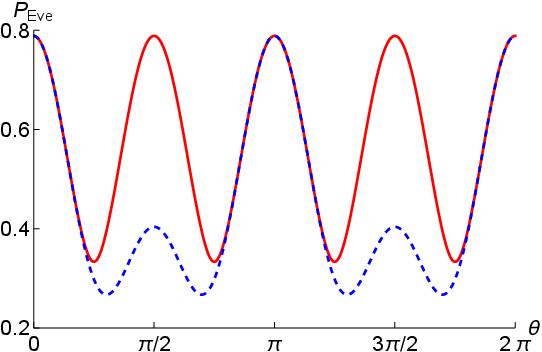}
\end{center}
\caption{Plots of the probabilities that Eve correctly guesses at a random bit that Alice sends.
The solid red curve represents the probability $P_{\mbox{\scriptsize Eve}}$
under a condition that Eve takes the optimum strategy depending on the noise parameter $\theta$.
The dashed blue curve represents the probability $\tilde{P}_{\mbox{\scriptsize Eve}}$
under a condition that Eve utilizes the parameters $\alpha=\alpha_{0}$ and $\beta=\beta_{0}$,
where $\alpha_{0}$ and $\beta_{0}$ are the optimized parameters for $\theta=0$.
The maximum and minimum values of $P_{\mbox{\scriptsize Eve}}$ are given by $(1/6)(3+\sqrt{3})\simeq 0.7887$ and $1/3$, respectively.
As $\theta$ changes from zero to $\pi/2$, $\tilde{P}_{\mbox{\scriptsize Eve}}$ becomes smaller than $P_{\mbox{\scriptsize Eve}}$ considerably.}
\label{Figure-02}
\end{figure}

Eve must adjust the values of $\alpha$ and $\beta$ depending on the value of $\theta$ and maximize $P_{\mbox{\scriptsize Eve}}$.
We plot $P_{\mbox{\scriptsize Eve}}$ as a function of $\theta$ on condition that Eve uses the optimum $\alpha$ and $\beta$
with a solid red curve in Fig.~\ref{Figure-02}.
When $\theta=0$, $P_{\mbox{\scriptsize Eve}}$ attains the maximum value $(1/6)(3+\sqrt{3})\simeq 0.7887$.
If we put $\theta=\pi/4$, $P_{\mbox{\scriptsize Eve}}$ reaches the minimum value $1/3$.

Here, we draw attention to the following facts.
The above Eve's optimization for $\alpha$ and $\beta$ is effective if Eve knows the value of $\theta$.
Thus, this strategy is useless practically
because Eve does not know $\theta$ at all and she cannot perform the optimization.
To avoid this trouble, we describe the optimum values of $\alpha$ and $\beta$ for $\theta=0$ as $\alpha_{0}$ and $\beta_{0}$, respectively,
and assume that Eve eavesdrops with these fixed parameters $\alpha_{0}$ and $\beta_{0}$.
In Fig.~\ref{Figure-02},
the dashed blue curve represents the probability $\tilde{P}_{\mbox{\scriptsize Eve}}$ that is obtained from Eq.~(\ref{P-Eve-intercept-resend-attack-1})
with substitutions of $\alpha_{0}$ and $\beta_{0}$ into $\alpha$ and $\beta$, respectively.
As $\theta$ changes from zero to $\pi/2$,
$\tilde{P}_{\mbox{\scriptsize Eve}}$ becomes smaller than $P_{\mbox{\scriptsize Eve}}$ considerably.

\begin{figure}
\begin{center}
\includegraphics{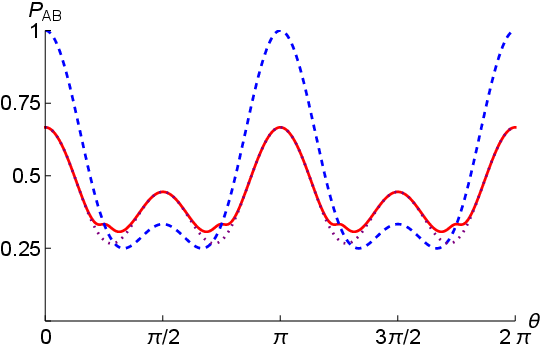}
\end{center}
\caption{Plots of the probabilities $P_{\mbox{\scriptsize AB}}$ that Alice and Bob share the same random bit correctly
on condition that both of their bases for encoding and observation correspond to each other as functions of the noise parameter $\theta$.
The solid red curve represents the probability when Eve makes the intercept/resend attack optimized for every $\theta$.
The dashed blue curve represents the probability that Eve does not mount the intercept/resend attack.
The dotted purple curve represents the probability on condition that Eve fixes the parameters $\alpha$ and $\beta$ at $\alpha_{0}$ and $\beta_{0}$ that are optimum for $\theta=0$.
For $0\leq\theta\leq\pi/4$, the probability without Eve's eavesdropping (the dashed blue curve) is larger than that with Eve's optimum attack (the solid red curve).
However, for $\pi/4\leq\theta\leq\pi/2$, the situation reverses itself.}
\label{Figure-03}
\end{figure}

Next, we estimate the probability $P_{\mbox{\scriptsize AB}}$ that Alice and Bob share the same random bit on condition that Eve eavesdrops on transmissions.
If Alice's basis for encoding and Bob's basis for observation correspond to each other,
the probability that Alice and Bob share a random bit is given by
\begin{eqnarray}
P_{\mbox{\scriptsize AB}}
&=&
\frac{1}{6}
[
\sum_{k=1,3,5}
P_{\mbox{\scriptsize Eve},k}
|\langle
\psi_{k}|V_{\mbox{\scriptsize Eve}}|\varphi_{01}\rangle|^{2}
+
\sum_{k=1,3,5}
(1-P_{\mbox{\scriptsize Eve},k})
|\langle\psi_{k}|V_{\mbox{\scriptsize Eve}}|\varphi_{10}\rangle|^{2} \nonumber \\
&&
+
\sum_{k=2,4,6}
P_{\mbox{\scriptsize Eve},k}
|\langle\psi_{k}|V_{\mbox{\scriptsize Eve}}|\varphi_{10}\rangle|^{2}
+
\sum_{k=2,4,6}
(1-P_{\mbox{\scriptsize Eve},k})
|\langle\psi_{k}|V_{\mbox{\scriptsize Eve}}|\varphi_{01}\rangle|^{2}
].
\end{eqnarray}
In Fig.~\ref{Figure-03}, we plot $P_{\mbox{\scriptsize AB}}$ as functions of the noise parameter $\theta$.
The solid red curve represents the probability if Eve mounts the optimized intercept/resend attack according to the noise parameter $\theta$.
The dashed blue curve represents the probability that Eve does not make any attacks.
The dotted purple curve represents the probability on condition that Eve attacks with fixed parameters $\alpha_{0}$ and $\beta_{0}$.
Looking at Fig.~\ref{Figure-03}, we note that the probability without Eve's attack (the dashed blue curve) is larger
than that with Eve's optimum attack (the solid red curve) for $0\leq\theta\leq\pi/4$.
However, this relationship is reversed for $\pi/4\leq\theta\leq\pi/2$.

The above fact implies the following.
In general, the probability that Alice and Bob share the same random bit via the noisy quantum channel is smaller than that via an ideal noiseless quantum channel.
Moreover, if Eve mounts the intercept/resend attack,
she disturbs the state transmitted and we can suppose that the probability that Alice and Bob share the same random bit becomes smaller.
Therefore,
we can expect that legitimate users can detect Eve's malicious acts by monitoring the probability.
However, Fig.~\ref{Figure-03} tells us that Eve's attack can let the probability be large in specific situations.
Hence, Alice and Bob cannot notice Eve's disturbance even if they monitor the probability that they succeed in sharing the same random bit.

The reversal of the relationship between the probabilities occurs when they are lower than $1/2$.
Thus, some might say that Alice and Bob can share the same random bit
if Bob inverses the value of his bit.
In this case, he obtain the same random bit with probability $1-P$ instead of $P$.
However, they cannot use this method practically because they do not know the value of $\theta$ and they cannot carry out statistical analyses for specific values of $\theta$.

\section{\label{section-entangling-probe-attack}Entangling probe attacks}
In this section, we consider an attack in which Eve lets her probe interact with the two qubits Alice sends,
keeps it on hand,
and observes it
after Alice and Bob disclose the bases for encoding and observation.
We name this scenario an entangling probe attack.
Because Eve's degree of freedom for this attack is very large, we can hardly analyse the security against it.
Thus, in this section, we focus on the following simplified entangling probe attack.

First, Eve applies the projection for detecting the parity bit to the state that Alice sends via the collective unitary noise channel and obtain
$|\tilde{\psi}_{j,t}\rangle$
for $j\in\{0,1\}$ and $t\in\{x,y,z\}$ in the form,
\begin{eqnarray}
|\tilde{\psi}_{j,t}\rangle
&=&
\Pi^{-}(U\otimes U)|j_{t}\rangle
\quad
\mbox{for $j\in\{0,1\}$ and $t\in\{x,y,z\}$},
\label{tilde-psi-j-t-0}
\end{eqnarray}
where
$|0_{x}\rangle=|\Psi^{+}\rangle$,
$|1_{x}\rangle=|\Psi^{-}\rangle$,
$|0_{y}\rangle=|i\Psi^{+}\rangle$,
$|1_{y}\rangle=|i\Psi^{-}\rangle$,
$|0_{z}\rangle=|01\rangle$,
$|1_{z}\rangle=|10\rangle$.
We show explicit forms of $|\tilde{\psi}_{j,t}\rangle$ in Appendix~\ref{appendix-tilde-psi-j-t}.
Second, Eve attaches an initialized auxiliary qubits (a probe) $|X\rangle$ to
$|\tilde{\psi}_{j,t}\rangle$.
Here, we draw attention to the fact that $|\tilde{\psi}_{j,t}\rangle$ lies on the two-dimensional Hilbert space spanned by $\{|0_{z}\rangle,|1_{z}\rangle\}$.

Third, Eve applies the following unitary transformation $U_{\mbox{\scriptsize Eve}}$ to $|\tilde{\psi}_{j,t}\rangle|X\rangle$:
\begin{eqnarray}
U_{\mbox{\scriptsize Eve}}|0_{z}\rangle|X\rangle
&=&
\sqrt{F}|0_{z}\rangle|A\rangle+\sqrt{1-F}|1_{z}\rangle|B\rangle, \nonumber \\
U_{\mbox{\scriptsize Eve}}|1_{z}\rangle|X\rangle
&=&
\sqrt{F'}|1_{z}\rangle|C\rangle+\sqrt{1-F'}|0_{z}\rangle|D\rangle,
\label{Eve-unitary-transformation-z}
\end{eqnarray}
where $|A\rangle$, $|B\rangle$, $|C\rangle$, and $|D\rangle$ are normalized arbitrary states.
Fourth, Eve leaves the probe at hand and sends the two qubits she steals to Bob via the ideal noiseless quantum channel.
Fifth, after Alice and Bob disclose the bases they use for encoding and observation through the classical channel in public,
Eve observes the probe according to Alice and Bob's public announcements.
As a result of this observation,
Eve guesses at the random bit shared by Alice and Bob.
In this attack, it is important that Eve can change the method for measuring her probe depending on public information Alice and Bob disclose through the classical channel.

The dimension of a Hilbert space for Eve's probe is equal to four at the most.
Here, we assume that Alice and Bob encode and observe the two-qubit state with $\sigma_{z}$ basis.
Eve applies $U_{\mbox{\scriptsize Eve}}$ and observes the probe.
On the one hand, we assume that she sends $|0_{z}\rangle$ to Bob if Eve detects $|A\rangle$.
Then, the probability that Alice and Bob share the same random bit is given by
$\langle\tilde{\psi}_{0,z}|\tilde{\psi}_{0,z}\rangle F=(\cos^{4}\theta+\sin^{4}\theta)F$.
On the other hand, we assume that Eve sends $|1_{z}\rangle$ to Bob if she detects $|C\rangle$.
Then, the probability that Alice and Bob share the same random bit is equal to
$\langle\tilde{\psi}_{1,z}|\tilde{\psi}_{1,z}\rangle F'=(\cos^{4}\theta+\sin^{4}\theta)F'$.
Alice and Bob can gather events in which they use the same basis for encoding and observation,
estimate the probabilities that they share the same random bit, and obtain $F$ and $F'$ statistically.
If $F$ is not equal to $F'$, Alice and Bob can notice Eve's disturbance.
Thus, Eve must set $F=F'$.
Here, we can put $1/2\leq F\leq 1$.

It is very difficult to find Eve's best strategy on condition that $U_{\mbox{\scriptsize Eve}}$ is an arbitrary unitary transformation.
Thus, we consider how to reduce degrees of freedom of $U_{\mbox{\scriptsize Eve}}$ by imposing symmetries on $U_{\mbox{\scriptsize Eve}}$.
Here, we draw attention to the following facts.
We suppose that Eve does not know the value of $\theta$ at all.
Thus, $U_{\mbox{\scriptsize Eve}}$ must not depend on $\theta$.

Bru{\ss} imposed the following symmetry on $U_{\mbox{\scriptsize Eve}}$ \cite{Bruss1998}:
\begin{eqnarray}
|\langle 0_{t}|U_{\mbox{\scriptsize Eve}}|0_{t}\rangle|X\rangle|^{2}
&=&
|\langle 1_{t}|U_{\mbox{\scriptsize Eve}}|1_{t}\rangle|X\rangle|^{2} \nonumber \\
&=&F
\quad
\mbox{for $t\in\{x,y,z\}$}.
\label{Bruss-symmetry-0}
\end{eqnarray}
From the above relationship, we obtain
\begin{equation}
\langle A|D\rangle
+
\langle B|C\rangle
=
0,
\label{equation-Bruss-1}
\end{equation}
\begin{equation}
\langle A|B\rangle
+
\langle D|C\rangle
=
0,
\label{equation-Bruss-2}
\end{equation}
\begin{equation}
\mbox{Re}\langle B|D\rangle = 0,
\label{equation-Bruss-3}
\end{equation}
\begin{equation}
\mbox{Re}\langle A|C\rangle
=
2-\frac{1}{F}.
\label{equation-Bruss-4}
\end{equation}

Further, according to Cirac and Gisin's work,
we apply the following symmetries to $U_{\mbox{\scriptsize Eve}}$ \cite{Cirac1997}.
For sake of simplicity, we adopt notations as
\begin{eqnarray}
U_{\mbox{\scriptsize Eve}}|0_{t}\rangle|X\rangle
&=&
|0_{t}\rangle|E^{t}_{00}\rangle
+
|1_{t}\rangle|E^{t}_{01}\rangle, \nonumber \\
U_{\mbox{\scriptsize Eve}}|1_{t}\rangle|X\rangle
&=&
|0_{t}\rangle|E^{t}_{10}\rangle
+
|1_{t}\rangle|E^{t}_{11}\rangle \quad \mbox{for $t\in\{x,y,z\}$}.
\label{unitary-transformation-qubit-probe}
\end{eqnarray}
We impose the following symmetries on Eve's attack:
\begin{equation}
\langle E^{x}_{ij}|E^{x}_{kl}\rangle
=
\langle E^{y}_{ij}|E^{y}_{kl}\rangle
=
\langle E^{z}_{ij}|E^{z}_{kl}\rangle
\quad
\mbox{for $i,j,k,l\in\{0,1\}$}.
\label{Eve-strategy-symmetric-1}
\end{equation}
This assumption implies that Eve's optimum strategy
forces Eve's unitary operator $U_{\mbox{\scriptsize Eve}}$
into acting on the transmitted qubits and the probe in the same way regardless of which basis Alice and Bob choose,
in other words, the $t$ basis $\forall t\in\{x,y,z\}$.

From Eqs.~(\ref{equation-Bruss-1}), (\ref{equation-Bruss-2}), (\ref{equation-Bruss-3}), (\ref{equation-Bruss-4}),
and (\ref{Eve-strategy-symmetric-1}),
we obtain
\begin{eqnarray}
\langle E^{t}_{00}|E^{t}_{00}\rangle
&=&
\langle E^{t}_{11}|E^{t}_{11}\rangle
=F, \nonumber \\
\langle E^{t}_{01}|E^{t}_{01}\rangle
&=&
\langle E^{t}_{10}|E^{t}_{10}\rangle
=1-F, \nonumber \\
\langle E^{t}_{00}|E^{t}_{01}\rangle
&=&
\langle E^{t}_{00}|E^{t}_{10}\rangle
=
\langle E^{t}_{01}|E^{t}_{10}\rangle \nonumber \\
&=&
\langle E^{t}_{01}|E^{t}_{11}\rangle
=
\langle E^{t}_{10}|E^{t}_{11}\rangle \nonumber \\
&=&
0, \nonumber \\
\langle E^{t}_{00}|E^{t}_{11}\rangle
&=&
F\cos\alpha
\quad
\mbox{for $t\in\{x,y,z\}$},
\label{Cirac-Gisin-relation-1}
\end{eqnarray}
\begin{equation}
\cos\alpha
=
2-\frac{1}{F},
\label{Cirac-Gisin-relation-2}
\end{equation}
where $0\leq\alpha\leq\pi/2$.
We show derivations of Eqs.~(\ref{Cirac-Gisin-relation-1}) and (\ref{Cirac-Gisin-relation-2})
in Supplemental Material \cite{Supplemental-Material}.
The relationships of Eqs.~(\ref{Cirac-Gisin-relation-1}) and (\ref{Cirac-Gisin-relation-2}) are the same as the results obtained by
Bechmann-Pasquinucci and Gisin
\cite{Bechmann-Pasquinucci1999}.

Here, we estimate the probability that Eve correctly guesses at the random bit Alice sends on condition that both Alice and Bob choose the $\sigma_{z}$ basis
for encoding and observation.
We give a name for the system of the state $|\tilde{\psi}_{j,z}\rangle$ for $j\in\{0,1\}$ that Eve steals via the quantum channel as Q.
We give the name of the system of Eve's probe as E.
Eve must distinguish between the following two density operators:
\begin{equation}
\tilde{\rho}^{z}_{j}
=
\mbox{Tr}_{\mbox{\scriptsize Q}}
(
U_{\mbox{\scriptsize Eve}}|\tilde{\psi}_{j,z}\rangle|X\rangle
\langle X|\langle\tilde{\psi}_{j,z}|U_{\mbox{\scriptsize Eve}}^{\dagger}
)
\quad \mbox{for $j\in\{0,1\}$}.
\end{equation}
We set an orthonormal basis
$\{|e_{1}\rangle,|e_{2}\rangle,|e_{3}\rangle,|e_{3}\rangle\}$ for the system E as
\begin{eqnarray}
|e_{1}\rangle
&=&
\frac{1}{\sqrt{F}}|E_{00}^{z}\rangle, \nonumber \\
|e_{2}\rangle
&=&
\frac{1}{\sqrt{F}\sin\alpha}(|E_{11}^{z}\rangle-\cos\alpha|E_{00}^{z}\rangle), \nonumber \\
|e_{3}\rangle
&=&
\frac{1}{\sqrt{1-F}}|E_{01}^{z}\rangle, \nonumber \\
|e_{4}\rangle
&=&
\frac{1}{\sqrt{1-F}}|E_{10}^{z}\rangle.
\label{orthonormal-basis-E-0}
\end{eqnarray}
Using Eq.~(\ref{orthonormal-basis-E-0}), we can write down
$\tilde{\rho}^{z}_{0}$ and $\tilde{\rho}^{z}_{1}$ as $4\times 4$ matrices.

The maximum value of the probability that Eve distinguishes between $\tilde{\rho}^{z}_{0}$ and $\tilde{\rho}^{z}_{1}$
is given as follows \cite{Fuchs1999}:
\begin{equation}
Q_{\mbox{\scriptsize Eve}}^{z,\mbox{\scriptsize max}}
=
\frac{1}{4}
[
||\tilde{\rho}^{z}_{0}-\tilde{\rho}^{z}_{1}||_{\mbox{\scriptsize t}}
+
\mbox{Tr}(\tilde{\rho}^{z}_{0})
+
\mbox{Tr}(\tilde{\rho}^{z}_{1})
],
\label{Q-Eve-z-max-0}
\end{equation}
where
$||X||_{\mbox{\scriptsize t}}
=
\mbox{tr}|X|$
and
$|X|=\sqrt{X^{\dagger}X}$ for an arbitrary operator $X$.
Similarly, we can compute $Q_{\mbox{\scriptsize Eve}}^{x,\mbox{\scriptsize max}}$ and
$Q_{\mbox{\scriptsize Eve}}^{y,\mbox{\scriptsize max}}$.
We give explicit mathematical expressions of
$Q_{\mbox{\scriptsize Eve}}^{t,\mbox{\scriptsize max}}$ for $t\in\{x,y,z\}$
in Appendix~\ref{appendix-Q-xyz-max}.

Finally, we obtain the probability that Eve correctly guesses at a random bit that Alice sends in the form,
\begin{equation}
Q_{\mbox{\scriptsize Eve}}
=
\frac{1}{3}
(
Q_{\mbox{\scriptsize Eve}}^{x,\mbox{\scriptsize max}}
+
Q_{\mbox{\scriptsize Eve}}^{y,\mbox{\scriptsize max}}
+
Q_{\mbox{\scriptsize Eve}}^{z,\mbox{\scriptsize max}}
).
\end{equation}

Next, we evaluate the probability that Alice and Bob share the same random bit.
If both Alice and Bob choose the $\sigma_{z}$ basis for encoding and observation,
we obtain the probability as
\begin{eqnarray}
Q^{z}_{\mbox{\scriptsize AB}}
&=&
\frac{1}{2}
[
\mbox{Tr}(\tilde{\rho}^{z}_{0})
+
\mbox{Tr}(\tilde{\rho}^{z}_{1})
] \nonumber \\
&=&
F\cos^{2}\theta+(1-F)\sin^{4}\theta.
\end{eqnarray}
Similarly, we can compute $Q^{x}_{\mbox{\scriptsize AB}}$ and $Q^{y}_{\mbox{\scriptsize AB}}$.
Finally, we attain
\begin{equation}
Q_{\mbox{\scriptsize AB}}
=
\frac{1}{3}
(
Q^{x}_{\mbox{\scriptsize AB}}
+
Q^{y}_{\mbox{\scriptsize AB}}
+
Q^{z}_{\mbox{\scriptsize AB}}
).
\end{equation}

\begin{figure}
\begin{center}
\includegraphics{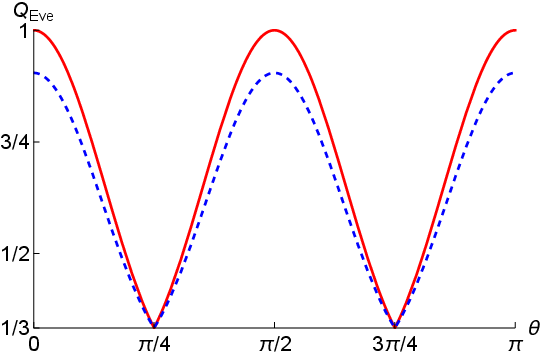}
\end{center}
\caption{Plots of $Q_{\mbox{\scriptsize Eve}}$ as functions of $\theta$ for $F=1/2$ and $F=3/4$.
The solid red and dashed blue curves represent $Q_{\mbox{\scriptsize Eve}}$
for $F=1/2$ and $F=3/4$, respectively.}
\label{Figure-04}
\end{figure}

\begin{figure}
\begin{center}
\includegraphics{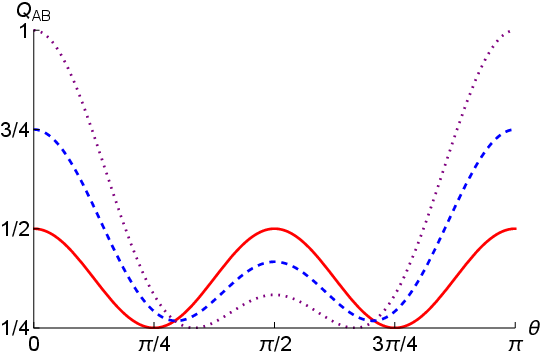}
\end{center}
\caption{Plots of $Q_{\mbox{\scriptsize AB}}$ as functions of $\theta$.
The solid red and dashed blue curves represent $Q_{\mbox{\scriptsize AB}}$ for $F=1/2$ and $F=3/4$, respectively.
The dotted purple curve represents the probability that Alice and Bob share the same random bit without Eve's eavesdropping.}
\label{Figure-05}
\end{figure}

The function of $Q_{\mbox{\scriptsize Eve}}$ depends on $\theta$ and $\alpha$.
If we optimize the value of $\alpha$ for an arbitrary $\theta$ so as to maximize $Q_{\mbox{\scriptsize Eve}}$,
we obtain $\alpha=\pi/2$, that is, $F=1/2$, from numerical calculations.
In Fig.~\ref{Figure-04}, we plot $Q_{\mbox{\scriptsize Eve}}$ as functions of $\theta$ for $F=1/2$ and $F=3/4$.
For both cases, $Q_{\mbox{\scriptsize Eve}}$ becomes minimum as $Q_{\mbox{\scriptsize Eve}}=1/3$ at $\theta=\pi/4$.

In Fig.~\ref{Figure-05}, we plot $Q_{\mbox{\scriptsize AB}}$ as functions of $\theta$ for $F=1/2$ and $F=3/4$.
For comparison, we draw a graph of the probability that Alice and Bob share the same random bit without Eve's eavesdropping.
Looking at Fig.~\ref{Figure-05}, we note that $Q_{\mbox{\scriptsize AB}}$ with Eve's disturbances is larger than that without Eve's eavesdropping
for $\theta\geq\theta_{0}\simeq 0.9210$.
The explicit value of $\theta_{0}$ is given by
\begin{equation}
\theta_{0}
=
\frac{1}{2}
\{
\pi
+
\arctan[
\frac{\sqrt{2+4(-2+\sqrt{3})}}{-2+\sqrt{3}}
]
\}.
\end{equation}

Looking at Figs.~\ref{Figure-04} and \ref{Figure-05},
we note the following facts.
For example, we assume that Eve chooses a strategy with $F=3/4$ for the noise parameter $\theta=3\pi/8$.
The probability $Q_{\mbox{\scriptsize AB}}$ for this case ($Q_{\mbox{\scriptsize AB}}=0.3196$, the dashed blue curve) is larger than that without Eve's eavesdropping
($Q_{\mbox{\scriptsize AB}}=0.2643$, the dotted purple curve).
Thus, in this case, Alice and Bob can hardly detect Eve's malicious acts.
However, in this case, $Q_{\mbox{\scriptsize Eve}}$ attains $0.6699$.
Thus, this scenario is very dangerous for Alice and Bob.

\section{\label{section-discussions}Discussions}
In this paper, we show that Alice and Bob cannot notice Eve's eavesdropping
in some specific cases for the fault-tolerant six-state QKD protocol over the collective unitary noise channel.
Thus, this protocol can be very dangerous for Alice and Bob.

As explained in Secs.~\ref{section-intercept-resend-attack}
and
\ref{section-entangling-probe-attack},
we can discover Eve's best strategies for the six-state protocol more easily than the BB84 scheme.
This is because the six-state protocol has more constraints than the BB84 scheme
so that the optimization problems for the six-state protocol become simpler than those of the BB84 scheme.

We do not examine whether or not the fault-tolerant BB84 scheme can be dangerous concerning the points indicated by our study.
This is a future subject.

In Sec,~\ref{section-entangling-probe-attack},
we consider the entangling probe attacks that do not depend on the noise parameter $\theta$.
We cannot answer the question of why Eve's optimized entangling probe attack does not depend on $\theta$.
There may be other good strategies for Eve that are adjustable depending on $\theta$.

In this paper, we estimate Alice, Bob, and Eve's probabilities,
for example,
$P_{\mbox{\scriptsize original}}$,
$P_{\mbox{\scriptsize noiseless subspace}}$,
$P_{\mbox{\scriptsize Eve}}$,
$\tilde{P}_{\mbox{\scriptsize Eve}}$,
$P_{\mbox{\scriptsize AB}}$,
$Q_{\mbox{\scriptsize Eve}}$,
and $Q_{\mbox{\scriptsize AB}}$,
on condition that Alice's basis for coding and Bob's basis for observation corresponds to each other.
The probability that Alice and Bob choose the same basis is equal to $1/3$.
By contrast, for the BB84 scheme, this probability is given by $1/2$.
Thus, the capacity of transmission for the six-state protocol is fewer than that for the BB84 scheme.

The six-state protocol has been proved to be unconditionally secure already \cite{Lo2001}.
However, from the point of view of practical use,
we think that our result is important.

\appendix

\section{\label{appendix-tilde-psi-j-t}
Explicit mathematical expressions of $|\tilde{\psi}_{j,t}\rangle$ given by Eq.~(\ref{tilde-psi-j-t-0})}
\begin{eqnarray}
|\tilde{\psi}_{0,x}\rangle
&=&
e^{i\delta}\cos(2\theta)|\Psi^{+}\rangle, \nonumber \\
|\tilde{\psi}_{1,x}\rangle
&=&
e^{i\delta}|\Psi^{-}\rangle, \nonumber \\
|\tilde{\psi}_{0,y}\rangle
&=&
e^{i\delta}
(\cos^{2}\theta|i\Psi^{+}\rangle-i\sin^{2}\theta|i\Psi^{-}\rangle), \nonumber \\
|\tilde{\psi}_{1,y}\rangle
&=&
e^{i\delta}
(\cos^{2}\theta|i\Psi^{-}\rangle+i\sin^{2}\theta|i\Psi^{+}\rangle), \nonumber \\
|\tilde{\psi}_{0,z}\rangle
&=&
e^{i\delta}
(\cos^{2}\theta|0_{z}\rangle-\sin^{2}\theta|1_{z}\rangle), \nonumber \\
|\tilde{\psi}_{1,z}\rangle
&=&
e^{i\delta}
(\cos^{2}\theta|1_{z}\rangle-\sin^{2}\theta|0_{z}\rangle),
\end{eqnarray}
\begin{eqnarray}
\langle\tilde{\psi}_{0,x}|\tilde{\psi}_{0,x}\rangle
&=&
\cos^{2}(2\theta), \nonumber \\
\langle\tilde{\psi}_{0,x}|\tilde{\psi}_{1,x}\rangle
&=&
0, \nonumber \\
\langle\tilde{\psi}_{1,x}|\tilde{\psi}_{1,x}\rangle
&=&
1, \nonumber \\
\langle\tilde{\psi}_{0,y}|\tilde{\psi}_{0,y}\rangle
&=&
\langle\tilde{\psi}_{1,y}|\tilde{\psi}_{1,y}\rangle
=
\cos^{4}\theta+\sin^{4}\theta, \nonumber \\
\langle\tilde{\psi}_{0,y}|\tilde{\psi}_{1,y}\rangle
&=&
2i\sin^{2}\theta\cos^{2}\theta,\nonumber \\
\langle\tilde{\psi}_{0,z}|\tilde{\psi}_{0,z}\rangle
&=&
\langle\tilde{\psi}_{1,z}|\tilde{\psi}_{1,z}\rangle
=
\cos^{4}\theta+\sin^{4}\theta, \nonumber \\
\langle\tilde{\psi}_{0,z}|\tilde{\psi}_{1,z}\rangle
&=&
-2\sin^{2}\theta\cos^{2}\theta.
\end{eqnarray}
From the above equations, we note that the phases $\phi$ and $\delta$ defined in Eq.~(\ref{collective-unitay-noise-1}) do not affect the improved protocol at all.

\section{\label{appendix-Q-xyz-max}Explicit mathematical expressions of $Q_{\mbox{\scriptsize Eve}}^{t,\mbox{\scriptsize max}}$ for $t\in\{x,y,z\}$
given by Eq.~(\ref{Q-Eve-z-max-0})}
\begin{eqnarray}
Q_{\mbox{\scriptsize Eve}}^{x,\mbox{\scriptsize max}}
&=&
\frac{1}{16}
\Biggl\{
12
+2 (2-F)\cos(4\theta) \nonumber \\
&&
+
F
\Bigl[
-6
+
\sqrt{
f(\theta,\alpha)
-2\sqrt{2}
g(\theta,\alpha)
\sin^{2}(2\theta)
} \nonumber \\
&&
+
\sqrt{
f(\theta,\alpha)
+2\sqrt{2}
g(\theta,\alpha)
\sin^{2}(2\theta)
}
\Bigr]
\Biggr\}, \nonumber \\
f(\theta,\alpha)
&=&
7
-8\cos(2\alpha)\cos^{2}(2\theta)
+\cos(8\theta), \nonumber \\
g(\theta,\alpha)
&=&
\sqrt{
11
+4\cos(4\theta)
-8\cos(2\alpha)[1+\cos(4\theta)]
+\cos(8\theta)
},
\end{eqnarray}
\begin{eqnarray}
Q_{\mbox{\scriptsize Eve}}^{y,\mbox{\scriptsize max}}
&=&
Q_{\mbox{\scriptsize Eve}}^{z,\mbox{\scriptsize max}} \nonumber \\
&=&
\frac{1}{8}
[
3
+\cos(4\theta)
+4|\cos(2\theta)|(1-F+F\sin\alpha)
],
\end{eqnarray}
where $0\leq\theta\leq\pi/2$ and $0\leq\alpha\leq\pi/2$.

\section*{Acknowledgements}
This work was supported by MEXT Quantum Leap Flagship Program (MEXT Q-LEAP) Grant Number JPMXS0120351339.

\end{document}


\title{Supplemental Material for ``Noise reduction caused by eavesdropping on six-state quantum key distribution over collective-noise channel''}

\author{Hiroo Azuma\thanks{On leave from Nisshin-scientia Co., Ltd.,
8F Omori Belport B, 6-26-2 MinamiOhi,
Shinagawa-ku, Tokyo 140-0013, Japan.
Email: hiroo.azuma@m3.dion.ne.jp, zuma@nii.ac.jp}
\\
\\
{\small Global Research Center for Quantum Information Science,}\\
{\small National Institute of Informatics,}\\
{\small 2-1-2 Hitotsubashi, Chiyoda-ku, Tokyo 101-8430, Japan}\\
}

\date{\today}

\maketitle

\setcounter{equation}{43}

\section*{Derivations of Eqs.~(31) and (32)}
From Eqs.~(23) with $F=F'$ and (29),
we obtain the following relations,
\begin{eqnarray}
|E^{z}_{00}\rangle
&=&
\sqrt{F}|A\rangle, \nonumber \\
|E^{z}_{01}\rangle
&=&
\sqrt{1-F}|B\rangle, \nonumber \\
|E^{z}_{10}\rangle
&=&
\sqrt{1-F}|D\rangle, \nonumber \\
|E^{z}_{11}\rangle
&=&
\sqrt{F}|C\rangle,
\label{unitary-transformation-z}
\end{eqnarray}
\begin{eqnarray}
|E^{x}_{00}\rangle
&=&
(1/2)[\sqrt{F}(|A\rangle+|C\rangle)+\sqrt{1-F}(|B\rangle+|D\rangle)], \nonumber \\
|E^{x}_{01}\rangle
&=&
(1/2)[\sqrt{F}(|A\rangle-|C\rangle)+\sqrt{1-F}(|D\rangle-|B\rangle)], \nonumber \\
|E^{x}_{10}\rangle
&=&
(1/2)[\sqrt{F}(|A\rangle-|C\rangle)+\sqrt{1-F}(|B\rangle-|D\rangle)], \nonumber \\
|E^{x}_{11}\rangle
&=&
(1/2)[\sqrt{F}(|A\rangle+|C\rangle)-\sqrt{1-F}(|B\rangle+|D\rangle)],
\label{unitary-transformation-x}
\end{eqnarray}
\begin{eqnarray}
|E^{y}_{00}\rangle
&=&
(1/2)[\sqrt{F}(|A\rangle+|C\rangle)+i\sqrt{1-F}(|D\rangle-|B\rangle)], \nonumber \\
|E^{y}_{01}\rangle
&=&
(1/2)[\sqrt{F}(|A\rangle-|C\rangle)+i\sqrt{1-F}(|D\rangle+|B\rangle)], \nonumber \\
|E^{y}_{10}\rangle
&=&
(1/2)[\sqrt{F}(|A\rangle-|C\rangle)-i\sqrt{1-F}(|B\rangle+|D\rangle)], \nonumber \\
|E^{y}_{11}\rangle
&=&
(1/2)[\sqrt{F}(|A\rangle+|C\rangle)+i\sqrt{1-F}(|B\rangle-|D\rangle)].
\label{unitary-transformation-y}
\end{eqnarray}

From Eqs.~(25), (26), (27), (28),
(\ref{unitary-transformation-z}), (\ref{unitary-transformation-x}), and (\ref{unitary-transformation-y}),
we obtain the following relations:
\begin{eqnarray}
\langle E^{t}_{00}|E^{t}_{00}\rangle
&=&
\langle E^{t}_{11}|E^{t}_{11}\rangle
=F, \nonumber \\
\langle E^{t}_{01}|E^{t}_{01}\rangle
&=&
\langle E^{t}_{10}|E^{t}_{10}\rangle
=1-F
\quad
\mbox{for $t\in\{x,y,z\}$}.
\end{eqnarray}

Moreover,
on the one hand,
Eq.~(\ref{unitary-transformation-z}) leads to
\begin{equation}
\langle E^{z}_{00}|E^{z}_{11}\rangle
=
F\langle A|C\rangle.
\label{z-00-11}
\end{equation}
On the other hand,
with the help of Eqs.~(27), (28), and (\ref{unitary-transformation-x}),
we obtain
\begin{equation}
\langle E^{x}_{00}|E^{x}_{11}\rangle
=
2F-1
-
(i/2)\sqrt{F(1-F)}[\mbox{Im}(\langle A|B\rangle-\langle D|C\rangle)+\mbox{Im}(\langle A|D\rangle-\langle B|C\rangle)].
\label{x-00-11}
\end{equation}
Substitution of Eqs.~(25), (26), (\ref{z-00-11}), and (\ref{x-00-11})
into $\langle E^{z}_{00}|E^{z}_{11}\rangle=\langle E^{x}_{00}|E^{x}_{11}\rangle$ gives
\begin{equation}
\mbox{Im}\langle A|C\rangle
=
-\sqrt{(1-F)/F}(\alpha+\beta),
\label{F-Im-AC-1}
\end{equation}
where
\begin{eqnarray}
\alpha
&=&
\mbox{Im}\langle A|B\rangle
=
-\mbox{Im}\langle D|C\rangle, \nonumber \\
\beta
&=&
\mbox{Im}\langle A|D\rangle
=
-\mbox{Im}\langle B|C\rangle.
\label{alpha-beta}
\end{eqnarray}
Further,
from Eqs.~(27), (28), and (\ref{unitary-transformation-y}),
we obtain
\begin{equation}
\langle E^{y}_{00}|E^{y}_{11}\rangle
=
2F-1
+
(i/2)\sqrt{F(1-F)}[\mbox{Re}(\langle A|B\rangle-\langle D|C\rangle)-\mbox{Re}(\langle A|D\rangle-\langle B|C\rangle)].
\label{y-00-11}
\end{equation}
Substitution of Eqs.~(25), (26), (\ref{z-00-11}), and (\ref{y-00-11})
into $\langle E^{z}_{00}|E^{z}_{11}\rangle=\langle E^{y}_{00}|E^{y}_{11}\rangle$ leads to
\begin{equation}
\mbox{Im}\langle A|C\rangle
=
\sqrt{(1-F)/F}(a-b),
\label{F-Im-AC-2}
\end{equation}
where
\begin{eqnarray}
a
&=&
\mbox{Re}\langle A|B\rangle
=
-\mbox{Re}\langle D|C\rangle, \nonumber \\
b
&=&
\mbox{Re}\langle A|D\rangle
=
-\mbox{Re}\langle B|C\rangle.
\label{a-b}
\end{eqnarray}
From Eqs.~(\ref{F-Im-AC-1}) and (\ref{F-Im-AC-2}),
we derive
\begin{equation}
-(\alpha+\beta)=a-b.
\label{alpha-beta-a-b}
\end{equation}

Because of Eqs.~(27) and (\ref{unitary-transformation-z}),
we obtain
\begin{equation}
\langle E^{z}_{01}|E^{z}_{10}\rangle = i(1-F)\mbox{Im}\langle B|D\rangle.
\label{z-01-10}
\end{equation}
Contrastingly, from Eqs.~(27), (28), (\ref{unitary-transformation-x}),
and (\ref{alpha-beta}),
we reach
\begin{equation}
\langle E^{x}_{01}|E^{x}_{10}\rangle = i\sqrt{F(1-F)}(\alpha-\beta).
\label{x-01-10}
\end{equation}
Thus,
with the help of $\langle E^{z}_{01}|E^{z}_{10}\rangle=\langle E^{x}_{01}|E^{x}_{10}\rangle$,
we arrive at
\begin{equation}
\sqrt{(1-F)/F}\mbox{Im}\langle B|D\rangle = \alpha - \beta.
\label{alpha-beta-2}
\end{equation}
Next,
Eqs.~(27), (28), (\ref{unitary-transformation-y}), and (\ref{a-b})
provide us with
\begin{equation}
\langle E^{y}_{01}|E^{y}_{10}\rangle = -i\sqrt{F(1-F)}(a+b).
\label{y-01-10}
\end{equation}
Thus,
$\langle E^{z}_{01}|E^{z}_{10}\rangle=\langle E^{y}_{01}|E^{y}_{10}\rangle$ leads to
\begin{equation}
-\sqrt{(1-F)/F}\mbox{Im}\langle B|D\rangle = a + b.
\label{a-b-2}
\end{equation}
Then,
from Eqs.~(\ref{alpha-beta-2}) and (\ref{a-b-2}),
we obtain
\begin{equation}
\alpha - \beta = -(a + b).
\label{alpha-beta-a-b-dash}
\end{equation}
From Eqs.~(\ref{alpha-beta-a-b}) and (\ref{alpha-beta-a-b-dash}),
we reach
\begin{equation}
\alpha = -a, \quad \beta = b.
\label{alpha-a-beta-b}
\end{equation}

Moreover,
from Eqs.~(\ref{unitary-transformation-z}), (\ref{alpha-beta}), and (\ref{a-b}),
we attain
\begin{equation}
\langle E^{z}_{00}|E^{z}_{01}\rangle
=
\sqrt{F(1-F)}(a+i\alpha).
\label{z-00-01}
\end{equation}
By contrast,
because of Eqs.~(26), (27), (\ref{unitary-transformation-x}),
(\ref{alpha-beta}), and (\ref{a-b}),
we obtain
\begin{equation}
\langle E^{x}_{00}|E^{x}_{01}\rangle
=
-(i/2)F\mbox{Im}\langle A|C\rangle+(i/2)(1-F)\mbox{Im}\langle B|D\rangle+\sqrt{F(1-F)}b.
\end{equation}
The relation $\langle E^{z}_{00}|E^{z}_{01}\rangle=\langle E^{x}_{00}|E^{x}_{01}\rangle$ leads to
\begin{equation}
a=b,
\label{a-b-3}
\end{equation}
\begin{equation}
\sqrt{F(1-F)}\alpha
=
-(1/2)F\mbox{Im}\langle A|C\rangle
+
(1/2)(1-F)\mbox{Im}\langle B|D\rangle.
\label{Im-AC}
\end{equation}
Further,
with the help of Eqs.~(26), (27), (\ref{unitary-transformation-y}),
(\ref{alpha-beta}), and (\ref{a-b}),
we arrive at
\begin{equation}
\langle E^{y}_{00}|E^{y}_{01}\rangle
=
-(i/2)F\mbox{Im}\langle A|C\rangle-(i/2)(1-F)\mbox{Im}\langle B|D\rangle-\sqrt{F(1-F)}\beta.
\end{equation}
From the relation $\langle E^{z}_{00}|E^{z}_{01}\rangle=\langle E^{y}_{00}|E^{y}_{01}\rangle$,
we obtain
\begin{equation}
a=-\beta,
\label{a-beta}
\end{equation}
\begin{equation}
\sqrt{F(1-F)}\alpha
=
-(1/2)F\mbox{Im}\langle A|C\rangle - (1/2)(1-F)\mbox{Im}\langle B|D\rangle.
\label{alpha-AC-BD}
\end{equation}

Because of Eqs.~(\ref{Im-AC}) and (\ref{alpha-AC-BD}),
we reach
\begin{equation}
\mbox{Im}\langle B|D\rangle = 0,
\label{Im-BD}
\end{equation}
so that we obtain $\langle B|D\rangle = 0$ from Eq.~(27).
Thus, with the help of Eq.~(\ref{z-01-10}), we attain $\langle E^{t}_{01}|E^{t}_{10}\rangle=0$ $\forall t\in\{x,y,z\}$.

Because of Eqs.~(\ref{a-b-2}) and (\ref{Im-BD}),
we reach
\begin{equation}
a+b=0.
\label{a-b-dash}
\end{equation}
Equations~(\ref{a-b-3}) and (\ref{a-b-dash}) lead to
\begin{equation}
a=b=0.
\label{a-b-0}
\end{equation}
Thus, from Eq.~(\ref{alpha-a-beta-b}), we derive
\begin{equation}
\alpha=\beta=0.
\label{alpha-beta-0}
\end{equation}

Hence, with the help of Eqs.~(\ref{Im-AC}), (\ref{Im-BD}), and (\ref{alpha-beta-0}),
we obtain
\begin{equation}
\mbox{Im}\langle A|C\rangle = 0.
\label{Im-AC-0}
\end{equation}
From Eq.~(\ref{z-00-01}), we reach $\langle E^{t}_{00}|E^{t}_{01}\rangle = 0$ $\forall t\in\{x,y,z\}$.

Because of the above discussion,
we obtain a relation,
\begin{equation}
\langle A|B\rangle
=
\langle D|C\rangle
=
\langle A|D\rangle
=
\langle B|C\rangle
=
0.
\label{ABCD-products}
\end{equation}
From the above results,
the following relations are automatically satisfied:
\begin{eqnarray}
\langle E^{t}_{00}|E^{t}_{10}\rangle
&=&
0, \nonumber \\
\langle E^{t}_{11}|E^{t}_{01}\rangle
&=&
0, \nonumber \\
\langle E^{t}_{11}|E^{t}_{10}\rangle
&=&
0
\quad\mbox{for $t\in\{x,y,z\}$}
\end{eqnarray}

Finally,
we conclude that Eve's unitary operator $U_{\mbox{\scriptsize Eve}}$ is uniquely and perfectly determined by a single parameter
$\mbox{Re}\langle A|C\rangle$ and the following relations hold:
\begin{equation}
F
=
\frac{1}{2-\mbox{Re}\langle A|C\rangle},
\label{fidelity-AC}
\end{equation}
\begin{equation}
\langle E^{t}_{00}|E^{t}_{11}\rangle = F\mbox{Re}\langle A|C\rangle \quad \forall t\in\{x,y,z\}.
\label{E00-E11-parameter}
\end{equation}

If we describe $\mbox{Re}\langle A|C\rangle=\cos\alpha$,
we obtain Eqs.~(31) and (32).